# Bound states of dispersion-managed solitons in a fiber laser at near zero dispersion

**L. M. Zhao, D. Y. Tang, and T. H. Cheng**

School of Electrical and Electronic Engineering, Nanyang Technological University, Singapore

**C. Lu and H. Y. Tam**

Department of Electronic and Information Engineering, Hong Kong Polytechnic University, Hong Kong

We report on the observation of various bound states of dispersion-managed (DM) solitons in a passively mode-locked Erbium-doped fiber ring laser at near zero net cavity group velocity dispersion (GVD). The generated DM solitons are characterized by their Gaussian-like spectral profile with no sidebands, which is distinct from those of the conventional solitons generated in fiber lasers with large net negative cavity GVD, of the parabolic pulses generated in fiber lasers with positive cavity GVD and negligible gain saturation and bandwidth limiting, and of the gain-guided solitons generated in fiber lasers with large positive cavity GVD. Furthermore, bound states of DM solitons with fixed soliton separations are also observed. We show that these bound solitons can function as a unit to form bound states themselves. Numerical simulations verified our experimental observations.

## 1. Introduction

Passively mode-locked Erbium-doped fiber lasers generally operate in the negative net cavity group velocity dispersion (GVD) regime, where after mode locking the pulse induced fiber nonlinear optical Kerr effect automatically balances the cavity dispersion [1-3]. Hence, solitary pulses are always formed. Furthermore, due to the cavity peak power clamping effect [4], multiple solitons are generated under strong pumping [5]. There have been shown three types of multiple soliton operation modes of the lasers: the soliton-bunching mode, where several soliton pulses bind and move together in the laser cavity; the stable soliton distribution mode, where soliton pulses are distributed randomly in the cavity with stable relative positions; and the stochastic soliton motion mode, where soliton pulses move randomly in the cavity. For the soliton-bunching mode, there are two sub-modes: loosely bound solitons, where the binding force between the solitons is weak. Therefore, the pulse separations can be easily altered by the environmental perturbations; and the multi-pulse solitons [6], where the solitons tightly bind together with discrete, fixed pulse separations. The multi-pulse solitons can even function as a unit, and form states of bound multi-pulse solitons [7].

The bound states of solitons formed in fiber lasers operating in the large normal or anomalous cavity dispersion regimes have been intensively investigated [6-13]. In a dispersion-managed fiber laser, the net cavity dispersion can be deliberately varied, and the laser can operate either in the conventional soliton, stretched-pulse [9-12, 14], self-similar parabolic pulse [13], or the gain-guided soliton [15] regime. For the solid-state lasers, Bob Proctor et al. have found that the optical spectra of the mode-locked pulses are essentially different as the total net cavity dispersion changed from -4fs$^2$ to +4fs$^2$ [16]. For the fiber lasers, Ph. Grelu et al. have studied the performance of soliton pairs in a stretched-pulse fiber laser operating in the normal path-averaged dispersion regime [10], and B. Ortac et al. have reported the generation of parabolic bound pulses in an Yb-doped fiber laser [13]. However, the operation of the mode-locked fiber lasers at near zero cavity GVD is so far not clearly addressed. In this paper, we report on the observation of various bound states of dispersion-managed (DM) solitons in an Erbium-doped fiber laser whose cavity GVD is around zero. We show that the optical spectra of the solitons formed in the laser have a smooth Gaussian-like profile without spectral sidebands, which is clearly different from those of the conventional laser solitons generated in fiber lasers with negative cavity GVD [1], of the stretched-pulses generated in dispersion-managed fiber lasers with positive cavity GVD [14], of the parabolic pulses generated in the dispersion-managed fiber lasers with positive cavity GVD, negligible gain saturation, and gain bandwidth limiting effects [13], or of the gain-guided solitons generated in fiber lasers with large positive cavity GVD [15]. Bound states of the solitons with fixed pulse separations have also been observed in the laser. Moreover, we show that the bound solitons can function as a unit to form other bound states of them. Numerical simulations have well confirmed our experimental observations.

## 2. Experimental studies

The fiber laser used is shown in Fig. 1. The cavity comprises a segment of 2.00 m Erbium-doped fiber (EDF) with GVD parameter of about -70 (ps/nm)/km sandwiched by two segments of standard single mode fiber (SMF) with GVD parameter of about 18 (ps/nm)/km, a 10% output coupler, a wavelength-division-multiplexer (WDM), and a fiber bench mounted with two sets of polarization controllers, one consists of two quarter-wave plates and the other two quarter-wave plates and one half-wave plate, a polarization dependent isolator, and a polarizer. The WDM and the output coupler are made of the standard SMF. The length of SMF is so selected that the total cavity dispersion is near zero. The nonlinear polarization rotation (NPR) technique is used to mode lock the laser.

It is well known that the cavity dispersion is wavelength dependent. However, the central wavelength of the mode-locked pulse shifts as the experimental condition is changed. In addition, the discrete bulk cavity components including the isolator and wave-plates also introduce slight dispersion, which needs to be considered in the case of near zero net cavity dispersion. In practice it is difficult to accurately measure the net cavity GVD under the various mode-locking states. To possibly allocate the net laser cavity dispersion near zero, we used a method as following: We first operate the laser in the conventional soliton regime. From the spectral sideband positions, we can roughly determine the total cavity GVD [17]. We then cut back either the length of the SMF or the EDF and repeated the above experiments. In this way, we could estimate the dispersion of the SMF and EDF used, as well as that of the bulk cavity components. The total dispersion of the isolator, the polarizer, and the wave-plates used in our laser is estimated as about $-8.4\times10^{-3}$ $ps^2$ at 1560nm wavelength, therefore, when the total length of the SMF is about 7.40 m, the net cavity dispersion is about $3.9\times10^{-4}$ $ps^2$.

At near zero cavity GVD, self-started mode-locking can be easily achieved. Fig. 2 shows for example the typical soliton spectrum and autocorrelation trace observed. Different from the spectrum of the conventional solitons, the optical spectrum of the mode-locked pulses of the laser has a Gaussian-like profile. In addition, no spectral sidebands are observed on the spectrum. The observed pulse spectrum resembles that of the DM solitons proposed by Haus et al [14]. For this reason we would call the mode-locked pulses the DM solitons.

Determined by the pump strength multiple DM solitons also appear in the laser. Similar to the conventional soliton operation of the fiber lasers [7-9] and the parabolic pulse operation in the positive net cavity dispersion regime [13], bound states of the DM solitons are also observed. In particular, depending on the detailed laser operation condition, we have observed bound solitons with various soliton number and soliton pulse separations. Figure 3 shows for example a state where two DM solitons bind. The separation between the solitons is 3.22ps, and the pulse width of each soliton is about 0.70 ps if the Gaussian profile is assumed. Based on the contour of the modulated spectrum, the spectral width of the DM soliton is estimated to be 22.7 nm. Thus, the time-bandwidth-product of the pulse is 1.96. Obviously it is a chirped soliton pulse. Fig. 4 shows further the cases of the bound solitons with soliton separations of 7.22ps (Fig. 4a) and 12.30ps (Fig. 4b).

Fig. 5a, Fig. 5b, and Fig. 5c, show cases where three, four, and five DM solitons are bound, respectively. From the autocorrelation traces it is to conclude that all the solitons in the bound states have identical pulse height and equal pulse separation, as the different peaks in the autocorrelation trace have the ratios of 1:2:3:2:1 for Fig. 5a, 1:2:3:4:3:2:1 for Fig. 5b, and 1:2:3:4:5:4:3:2:1 for Fig. 5c. The results shown in Fig. 5 were obtained when the orientations of the quarter-wave plates and the pump strength were changed. In our experiment with different orientations of the quarter-wave plates the linear cavity birefringence changes slightly. Therefore, the net phase delay between the two orthogonal cavity polarization components after one round-trip in the cavity is also changed [4]. We term this variable linear cavity phase delay the cavity phase delay bias. In the states shown in Fig. 5 the soliton pulse width is about 0.91 ps if the Gaussian profile is assumed. We note that although the number of pulse in each state is different, the soliton separations are fixed at 2.73 ps. Monitored by an autocorrelator with 50 ps maximum

scan range and a 50 GHz high-speed sampling oscilloscope (Agilent 86100A), we also confirmed experimentally that there is only one soliton bunch propagating in the cavity.

All these bound states of DM solitons are stable in the sense that after being obtained they can last several hours if no perturbation is introduced. Experimentally, bound states of the bound DM solitons were also observed as shown in Fig. 6. In the state at first every two DM solitons bind together forming a bound group, then the bound group functions as a unit and bind further with other bound group of the same type. The pulse-to-pulse separation in the primary bound solitons is still fixed at 2.73 ps. However, the soliton spacing in the secondary bound states is variable, In Fig. 6a it is 9.37 ps, and in Fig. 6b it is 8.00ps.

**3. Numerical simulations**

To confirm the formation of DM solitons and their bound states at the near zero cavity group velocity dispersion, we further numerically simulate the laser operation. We used exactly the same model and simulation techniques as described in Ref. 8 for the simulations. To be comparable with the experimental results, we used the following parameters for the current simulations: $\gamma$ = 3 $W^{-1}km^{-1}$; $k''$ = -0.023 $ps^2/m$ (SMF); $k''$ = 0.090 $ps^2/m$ (EDF); $k'''$ = $-1.29\times10^{-4}$ $ps^3/m$; $\Omega_g$ = 25 nm; gain saturation intensity $P_{sat}$=1000pJ; cavity length L = $2.4_{SMF}+2.0_{EDF}+5.0_{SMF}$ =9.4 m; cavity beat length $L_b$=L/2. The polarizer orientation to the fiber fast axis is $\psi = 0.125\pi$. The linear cavity phase delay bias is chosen as $1.4\pi$. Thus the net cavity GVD is about 0.0098 $ps^2$ at 1560nm. As shown by the numerical results, DM solitons and their bound state of solitons can be formed in a dispersion-managed laser even with net positive near zero cavity GVD. This result is in agreement of that of DM solitons formed in the dispersion-managed transmission line [18].

Fig. 7 shows the calculated mode-locked pulse and its spectrum. The temporal profile is closer to the Gaussian fitting than the $Sech^2$ fitting. Similar results are also obtained in lasers with net negative near zero cavity GVD. Obviously, when the net laser cavity GVD is near zero, the mode-locked pulse could have a very different optical spectrum to those of the conventional soliton pulse. Like the experimental observation, there are no sidebands on the optical spectrum. We believe this result could be a consequence of the nearly equal pulse broadening and compression experienced by the mode-locked pulse in one cavity roundtrip, because the net cavity GVD is near zero [14]. Numerically, we confirmed that the multiple pulse formation in the laser is still due to the cavity peak clamping effect. As a result of different linear cavity phase delay bias settings, the mode-locked pulses are clamped at different peak powers. The soliton effect of the pulses further determines their pulse width. Fig. 8 shows the bound state of two DM solitons when the small signal gain coefficient $g_0$=550. In obtaining the bound state we have used an arbitrary weak pulse as the initial state and let it circulate in the cavity. After several thousands of cavity round trips the laser output stabilized at the state shown. Once the state is obtained, further slightly changes the pump strength does not alter the pulse separation but their strength, indicating that they are indeed a stable state of the laser emission. Numerically we have calculated the phase difference between the two pulse peaks, it is about 12.5π. We note that A. Maruta *et al.* numerically studied bi-soliton formation and propagation in dispersion-managed systems described by the nonlinear Schrödinger equation (NLE) [19]. They have shown the existence of either in-phase or anti-phase bi-solitons in the systems. Our simulations further showed that stable bi-solitons with other phase differences are also possible in dispersion-managed lasers. As a laser is essentially a Ginzburg-Landau equation (GLE) system, and mode locking mechanism of a laser has the function of phase locking the pulses formed in the cavity, it is understandable of the phase difference. Depending on the cavity parameter settings and initial pulse state bound states of the solitons with other soliton separations have also been obtained in our simulations.

Numerically we also achieved the bound states of bound solitons as shown in Fig. 9. The parameters used are the same as above except that a different initial state and bigger $g_0$ are used to perform simulation. Fig. 9a shows that the soliton separation between the primary bound solitons is about 10 ps, while that in Fig. 9b is 20 ps. The pulse-to-pulse separations in both the primary bound solitons are about 3 ps in Fig. 9a and Fig. 9b. Further numerical simulations show that the pulse intensity of the individual pulse in the bound states can change with the pump power. Bigger pump power leads to higher pulse intensity. However, the pulse-to-pulse separation remains fixed under the change of pump power. Same as the single-pulse DM solitons, the primary bound DM solitons could also function as a unit to form bound states. Our simulation results clearly show that the primary bound DM solitons could be another intrinsic form of pulsation in the laser. Namely, it is a twin-pulse DM soliton.

Although the optical spectrum of the DM soliton formed in the laser has different appearance to those of the conventional laser solitons, which suggests that they could have different pulse shaping mechanisms, our studies show that the DM solitons still have exactly the same features as those of the conventional solitons. First, the multiple DM solitons are generated as a result of the cavity pulse peak clamping effect. Second, the multiple DM solitons interact with each other in the cavity and their interaction strength varies with their separation. When they interact directly, they can bind strongly together to form the so-called twin-pulse or multipulse DM solitons with fixed, discrete soliton separations. Furthermore, our studies also show that even when interaction between the solitons is weak, they can still form bound states of solitons, however, due to the perturbation of other effects, such as interaction mediated through the dispersive waves, their soliton separation has no longer fixed discrete values.

## 4. Conclusions

In conclusion, we have studied the operation of a DM soliton fiber laser at near zero cavity GVD. We have shown experimentally that the soliton spectrum of the laser has a Gaussian-like profile without spectral sidebands. Various bound states of the solitons either with different soliton number or soliton separations were observed. These bound states of solitons are characterized by that they have discrete fixed pulse separations once they are formed in the lasers. In particular, bound states of the bound solitons are also experimentally obtained, which demonstrates that the bound solitons could function as a unit, therefore, they could be regarded as a new form of the multi-pulse DM solitons. Numerical simulations have well confirmed our experimental observations.

## References:

1. D. J. Richardson, R. I. Laming, D. N. Payne, M. W. Philips, and V. J. Matsas, “320 fs soliton generation with passively mode-locked Erbium fibre laser,” Electron. Lett. 27, 730-732 (1991).
2. I. N. Duling III, “Subpicosecond all-fibre Erbium laser,” Electron. Lett. 27, 544-545 (1991).
3. M. Nakazawa, E. Yoshida, and Y. Kimura, “Low threshold, 290 fs Erbium-doped fiber laser with a nonlinear amplifying loop mirror pumped by InGaAsP laser diodes,” Appl. Phys. Lett. 59, 2073–2075 (1991).
4. D. Y. Tang, L. M. Zhao, B. Zhao, and A. Q. Liu, “Mechanism of multisoliton formation and soliton energy quantization in passively mode-locked fiber lasers,” Phys. Rev. A 72, 043816 (2005).
5. D. J. Richardson, R. I. Laming, D. N. Payne, V. J. Matsas, and M. W. Philips, “Pulse repetition rates in passive, self starting, femtosecond soliton fibre laser,” Electron. Lett. 27, 1451 (1991).
6. D. Y. Tang, W. S. Man, H. Y. Tam, and P. D. Drummond, “Observation of bound states of solitons in a passively mode-locked fiber laser”, Phys. Rev. A. 64, 033814 (2001).
7. D. Y. Tang, L. M. Zhao, and B. Zhao, “Multipulse bound solitons with fixed pulse separations formed by direct soliton interaction,” Appl. Phys. B 80, 239-242 (2005).
8. D. Y. Tang, B. Zhao, D. Y. Shen, C. Lu, W. S. Man, and H. Y. Tam, “Bound-soliton fiber laser,” Phys. Rev. A 66, 033806 (2002).
9. Ph. Grelu, F. Belhache, F. Gutty, and J. M. Soto-Crespo, “Phase-locked soliton pairs in a stretched-pulse fiber laser,” Opt. Lett. 27, 966 (2002).

10. Ph. Grelu, J. Béal, and J. M. Soto-Crespo, “Soliton pairs in a fiber laser: from anomalous to normal average dispersion regime,” Opt. Express 11, 2238-2243 (2003).
11. J. M. Soto-Crespo, N. Akhmediev, Ph. Grelu, and F. Belhache, “Quantized separations of phase-locked soliton pairs in fiber lasers,” Opt. Lett. 28, 1757-1759 (2003).
12. B. Ortac, A. Hideur, T. Chartier, M. Brunel, Ph. Grelu, H. Leblond, and F. Sanchez, “Generation of bound states of three ultra-short pulses with a passively mode-locked high-power Yb-doped double-clad fiber laser,” IEEE Photon. Technol. Lett. 16, 1274-1276 (2004).
13. B. Ortac, A. Hideur, M. Brunel, C. Chédot, J. Limpert, A. Tünnermann, and F. Ő. Ilday, “Generation of parabolic bound pulses from a Yb-fiber laser,” Opt. Express 6075-6083 (2006).
14. H. A. Haus, K. Tamura, L. E. Nelson, and E. P. Ippen, “Stretched-pulse additive-pulse mode-locking in fiber ring lasers: theory and experiment,” IEEE J. Quantum. Electron. 31, 591-598 (1995).
15. L. M. Zhao, D. Y. Tang, T. H. Cheng, and C. Lu, “Gain-guided solitons in dispersion-managed fiber lasers with large net cavity dispersion,” Opt. Lett. 31, 2957-2959 (2006).
16. Bob Proctor, Erik Westwig, and Frank Wise, “Characterization of a Kerr-lens mode-locked Ti:sapphire laser with positive group-velocity dispersion,” Opt. Lett. 18, 1654-1656 (1993).
17. M. L. Dennis, I. N. Duling III, “Intracavity dispersion measurement in modelocked fibre laser,” Electron. Lett. 29, 409-411 (1993).
18. J. H. B. Nijhof, N. J. Doran, W. Forysiak, and F. M. Knox, “Stable soliton-like propagation in dispersion managed systems with net anomalous, zero and normal dispersion,” Electron. Lett. 33, 1726-1727 (1997).

19. A. Maruta, T. Inoue, Y. Nonaka, and Y. Yoshika, “Bisoliton propagating in dispersion-managed system and its application to high-speed and long-haul optical transmission,” IEEE J. Sel. Top. Quantum Electron. 8, 640-650 (2002).

## Figure Captions:

Fig. 1 Experiment setup. $\lambda/4$ : quarter-wave plate; $\lambda/2$ : half-wave plate; PI: polarization-dependent isolator; P: polarizer; WDM: wavelength-division multiplexer; EDF: Erbium doped fiber.

Fig. 2 A typical dispersion-managed soliton. (a) autocorrelation trace; (b) optical spectrum.

Fig. 3 Bound state of two dispersion-managed solitons. (a) autocorrelation trace; (b) optical spectrum.

Fig. 4 Other bound states of two dispersion-managed solitons with different soliton separations of (a) 7.22 ps; (b) 12.30 ps.

Fig. 5 Multi-pulse dispersion-managed solitons (a) three-pulse; (b) four-pulse; (c) five-pulse.

Fig. 6 Bound states of twin-pulse dispersion-managed solitons with different soliton separations of (a) 9.37 ps; (b) 8.00 ps.

Fig. 7 Numerically calculated dispersion-managed solitons of the laser at $g_0$=550 (a) temporal profile; (b) optical spectrum.

Fig. 8 Numerically simulated bound state of two dispersion-managed solitons of the laser at $g_0$=550 (a) temporal profile; (b) optical spectrum.

Fig. 9 Numerically calculated bound states of twin-pulse dispersion-managed solitons with different soliton separations of (a) 10 ps; (b) 20 ps.

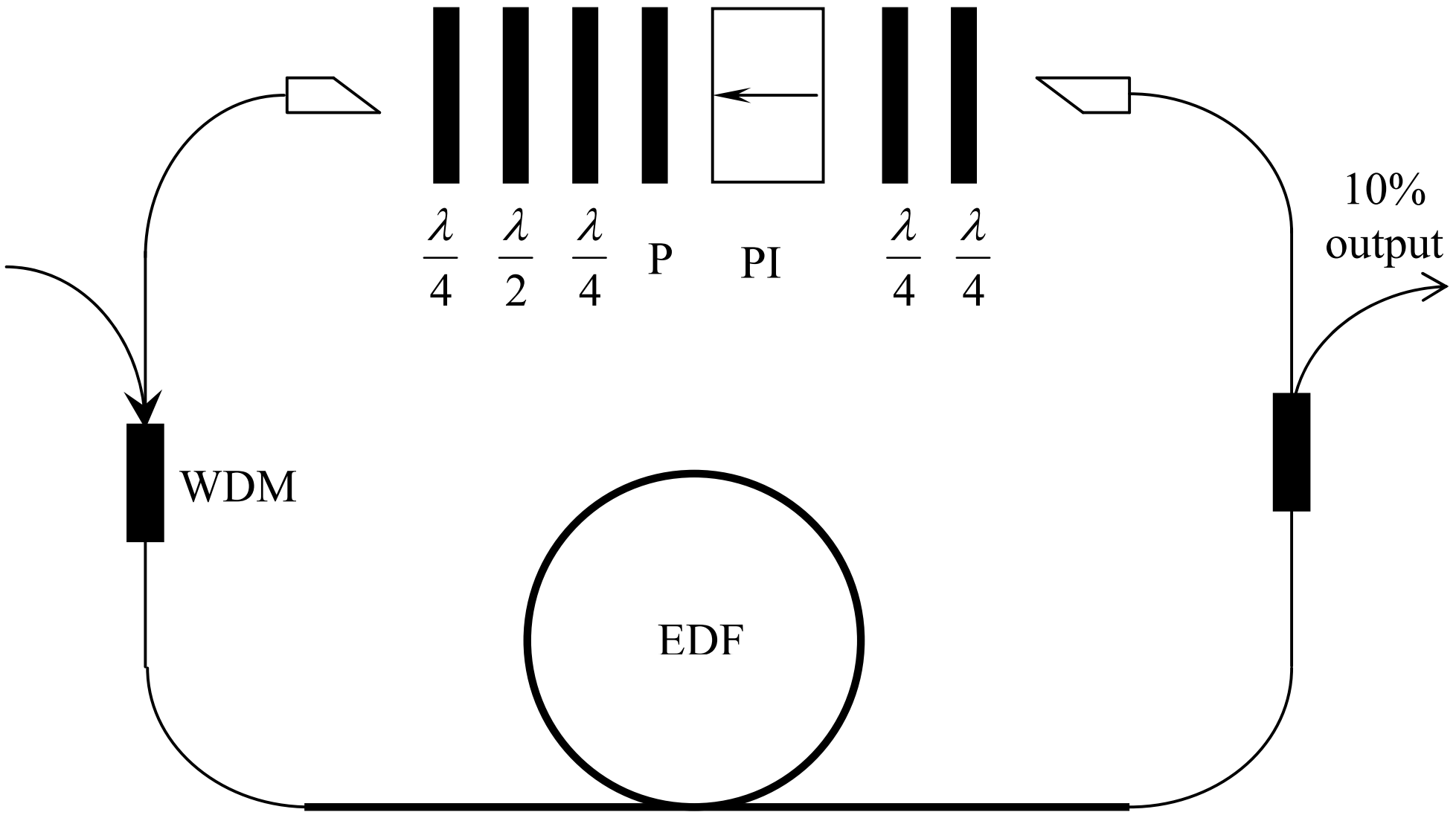

Fig. 1 L. M. Zhao et al. "Bound states of dispersion-managed solitons in a fiber laser with around zero dispersion"

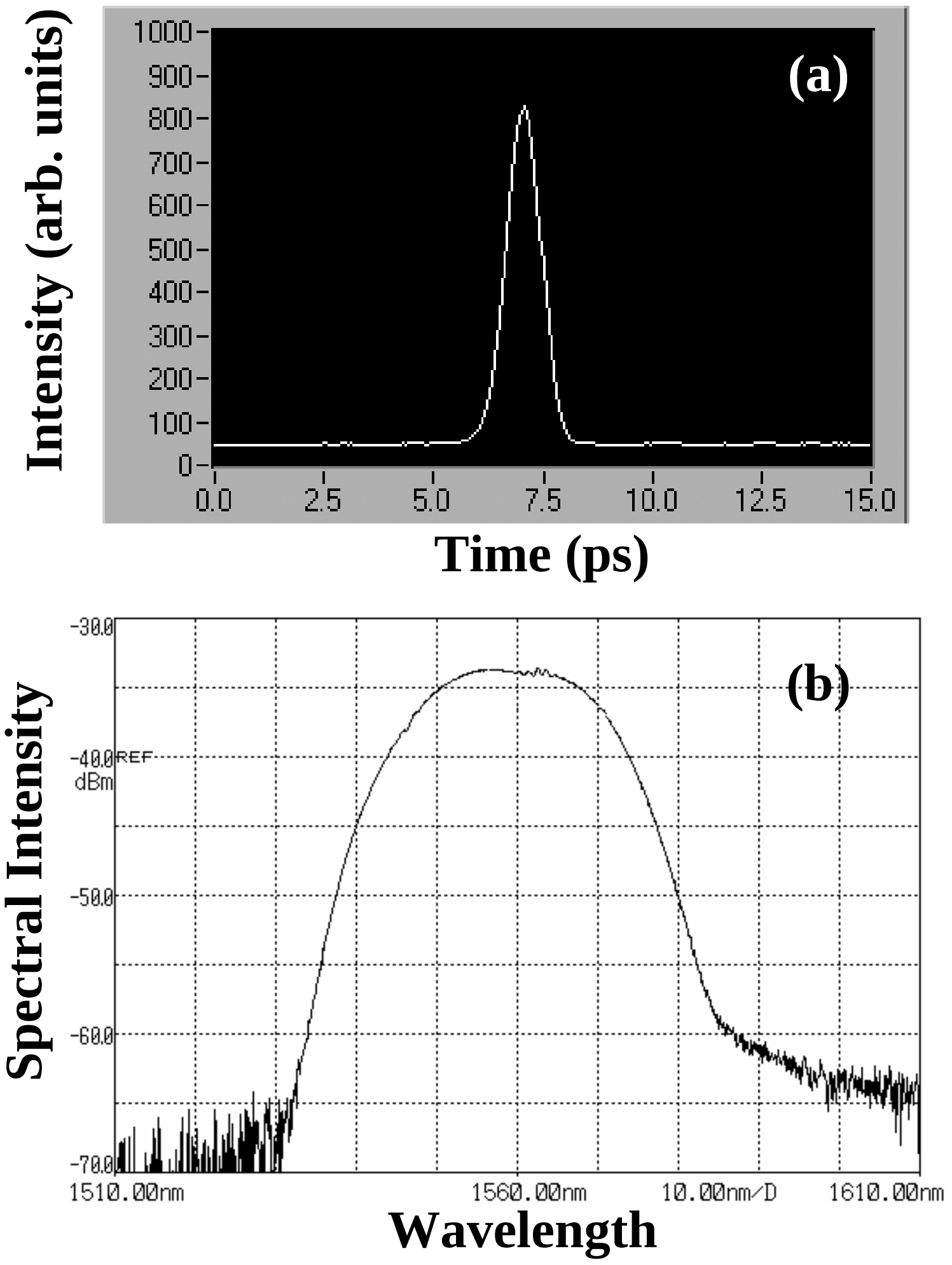

Fig. 2 L. M. Zhao et al. "Bound states of dispersion-managed solitons in a fiber laser with around zero dispersion"

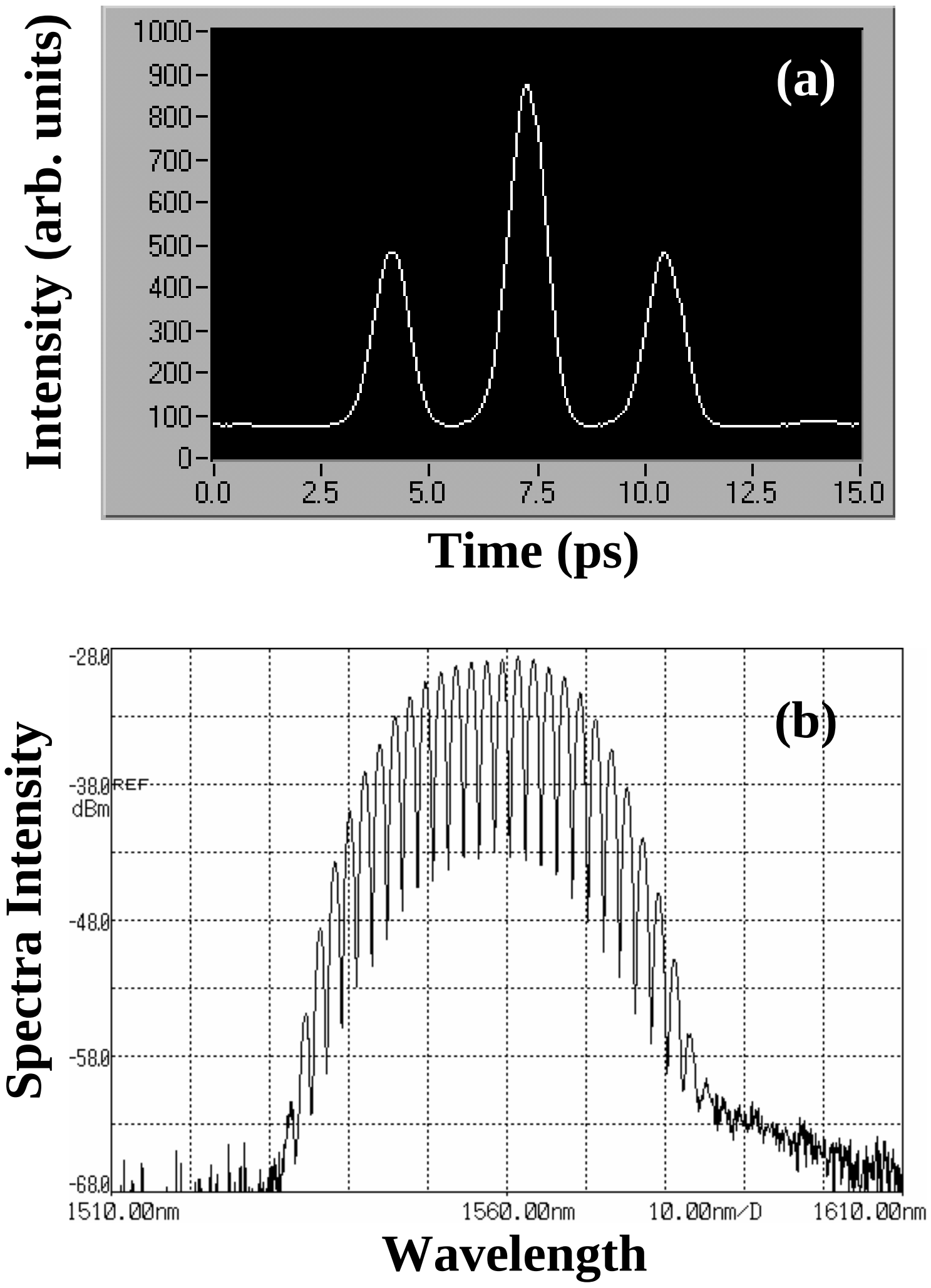

Fig. 3 L. M. Zhao et al. “Bound states of dispersion-managed solitons in a fiber laser with around zero dispersion”

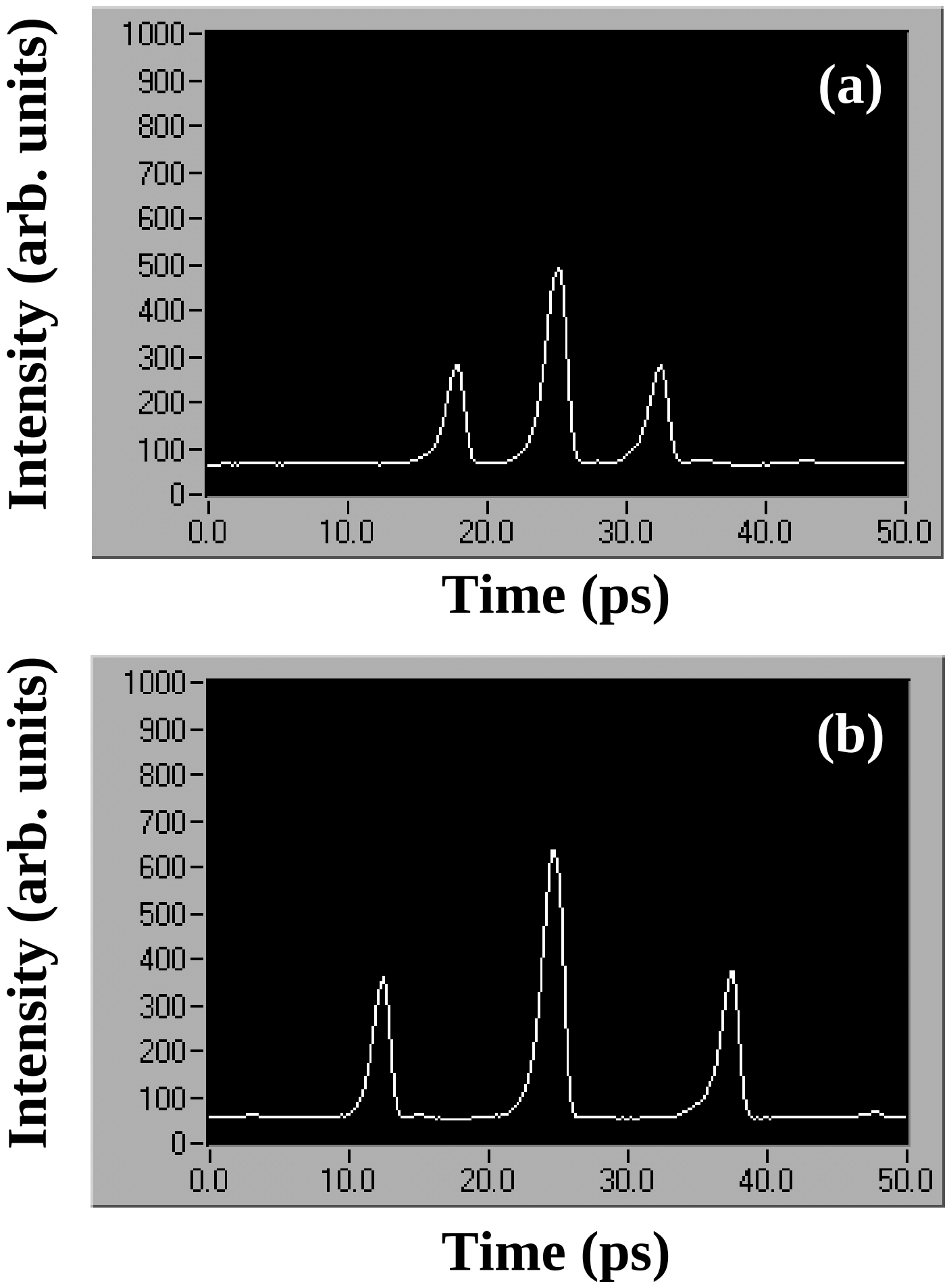

Fig. 4 L. M. Zhao et al. "Bound states of dispersion-managed solitons in a fiber laser with around zero dispersion"

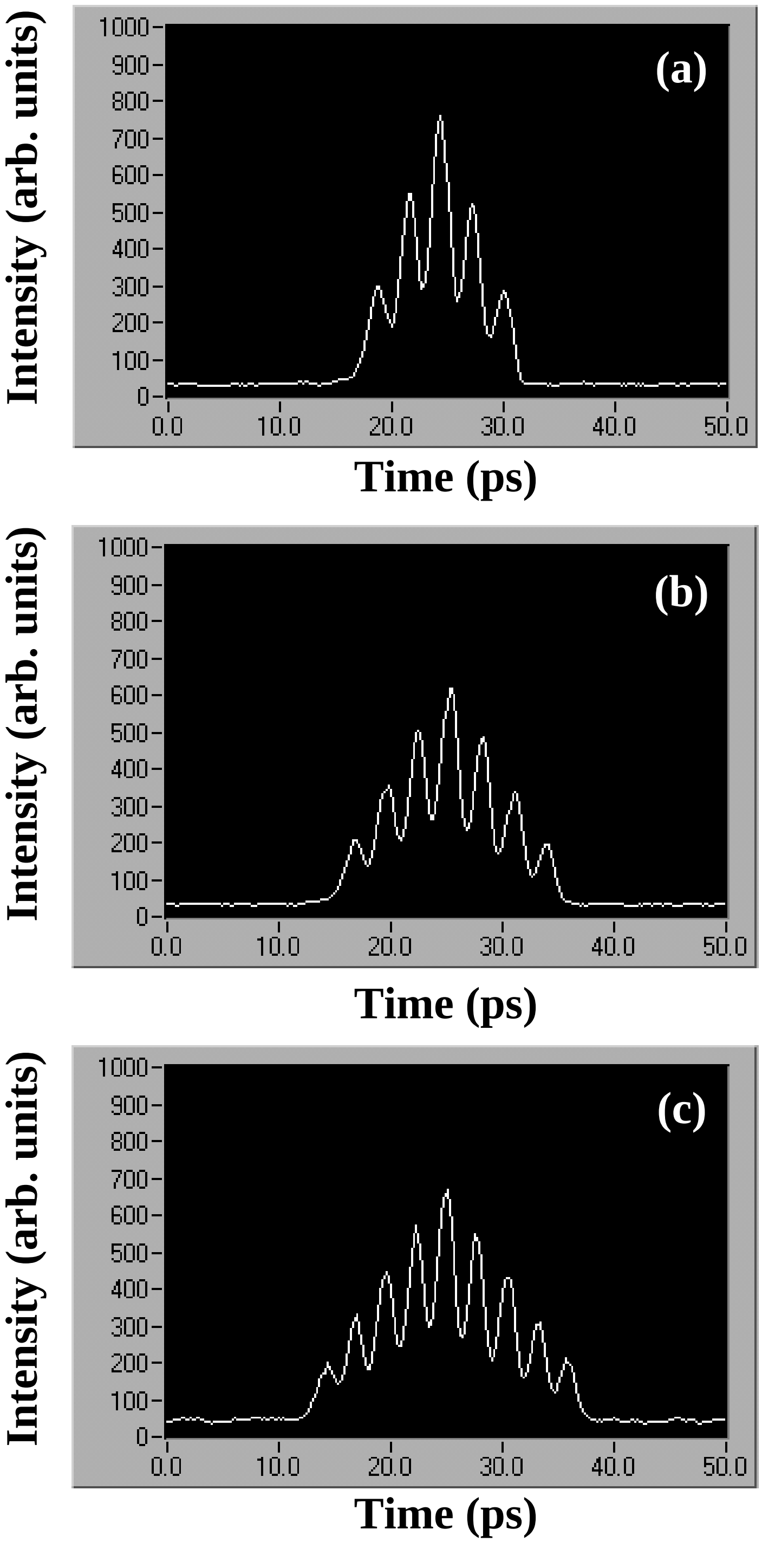

Fig. 5 L. M. Zhao et al. "Bound states of dispersion-managed solitons in a fiber laser with around zero dispersion"

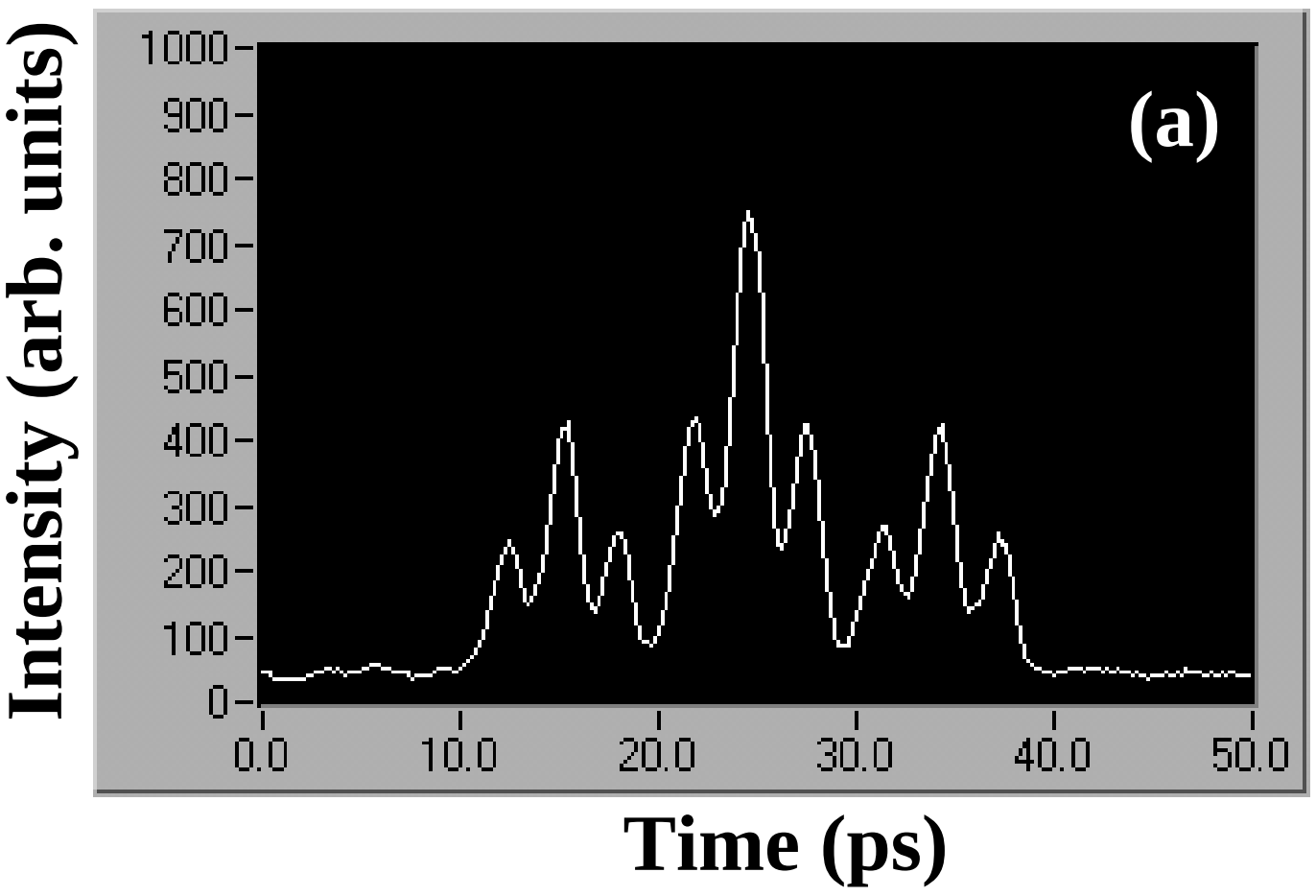

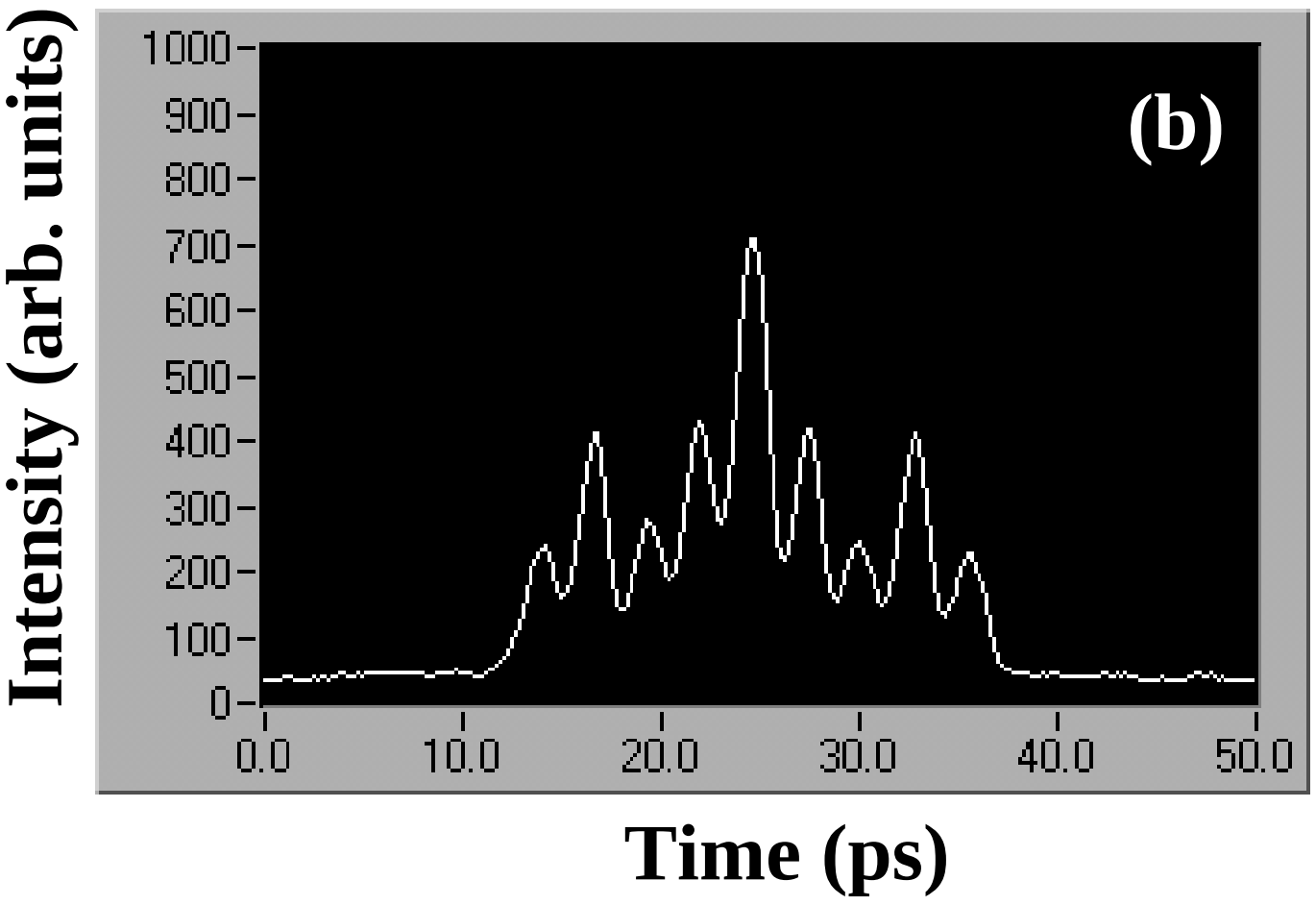

Fig. 6 L. M. Zhao et al. "Bound states of dispersion-managed solitons in a fiber laser with around zero dispersion"

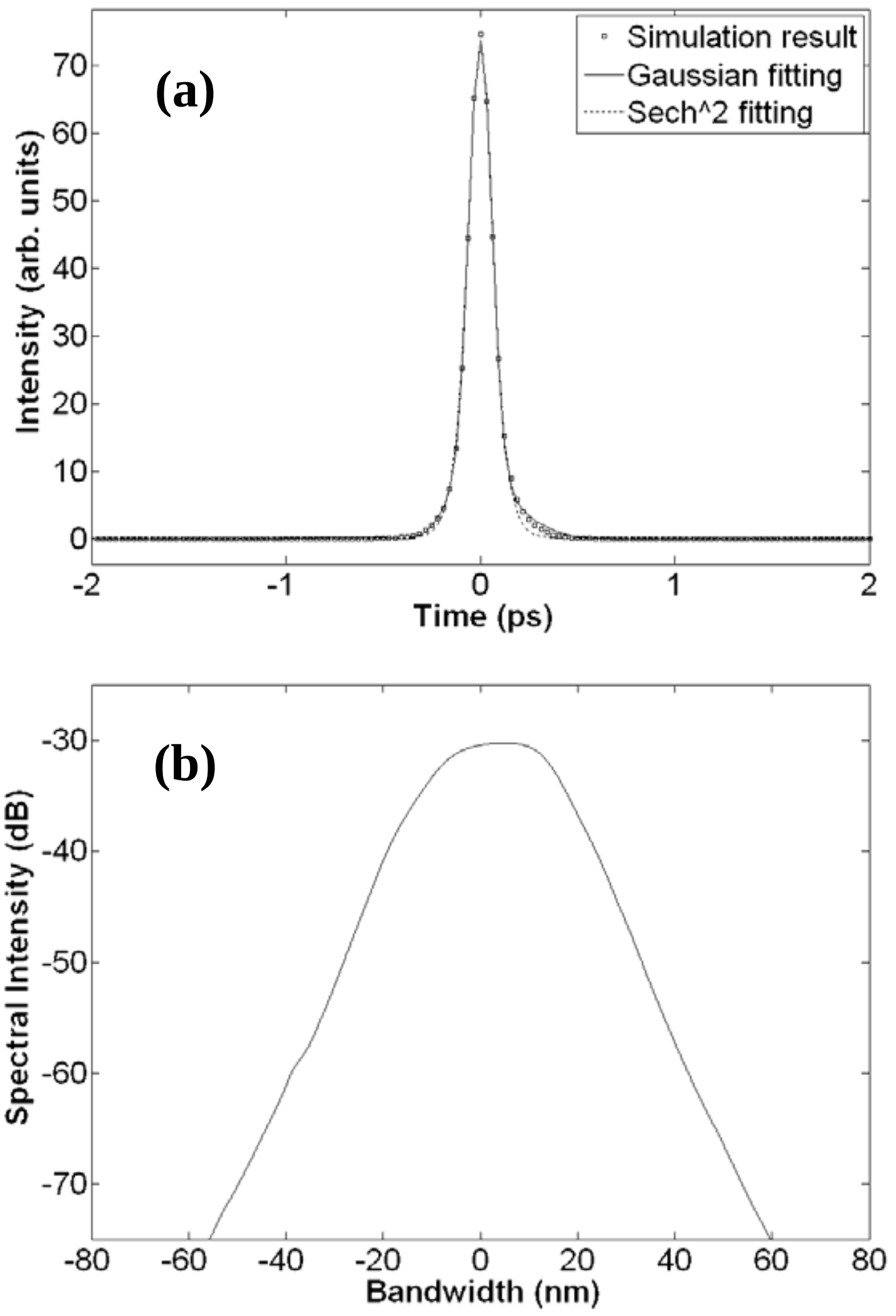

Fig. 7 L. M. Zhao et al. "Bound states of dispersion-managed solitons in a fiber laser with around zero dispersion"

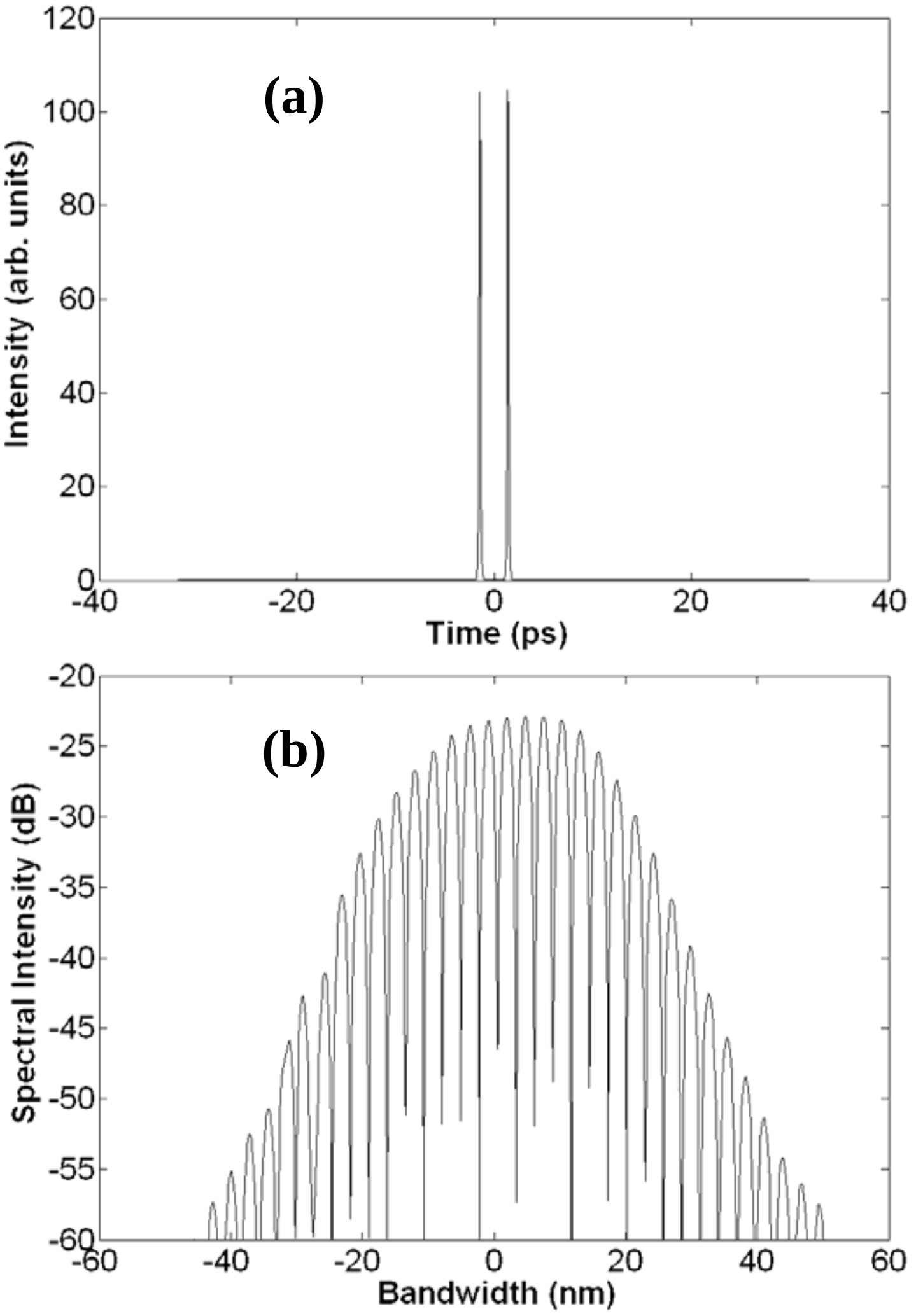

Fig. 8 L. M. Zhao et al. "Bound states of dispersion-managed solitons in a fiber laser with around zero dispersion"

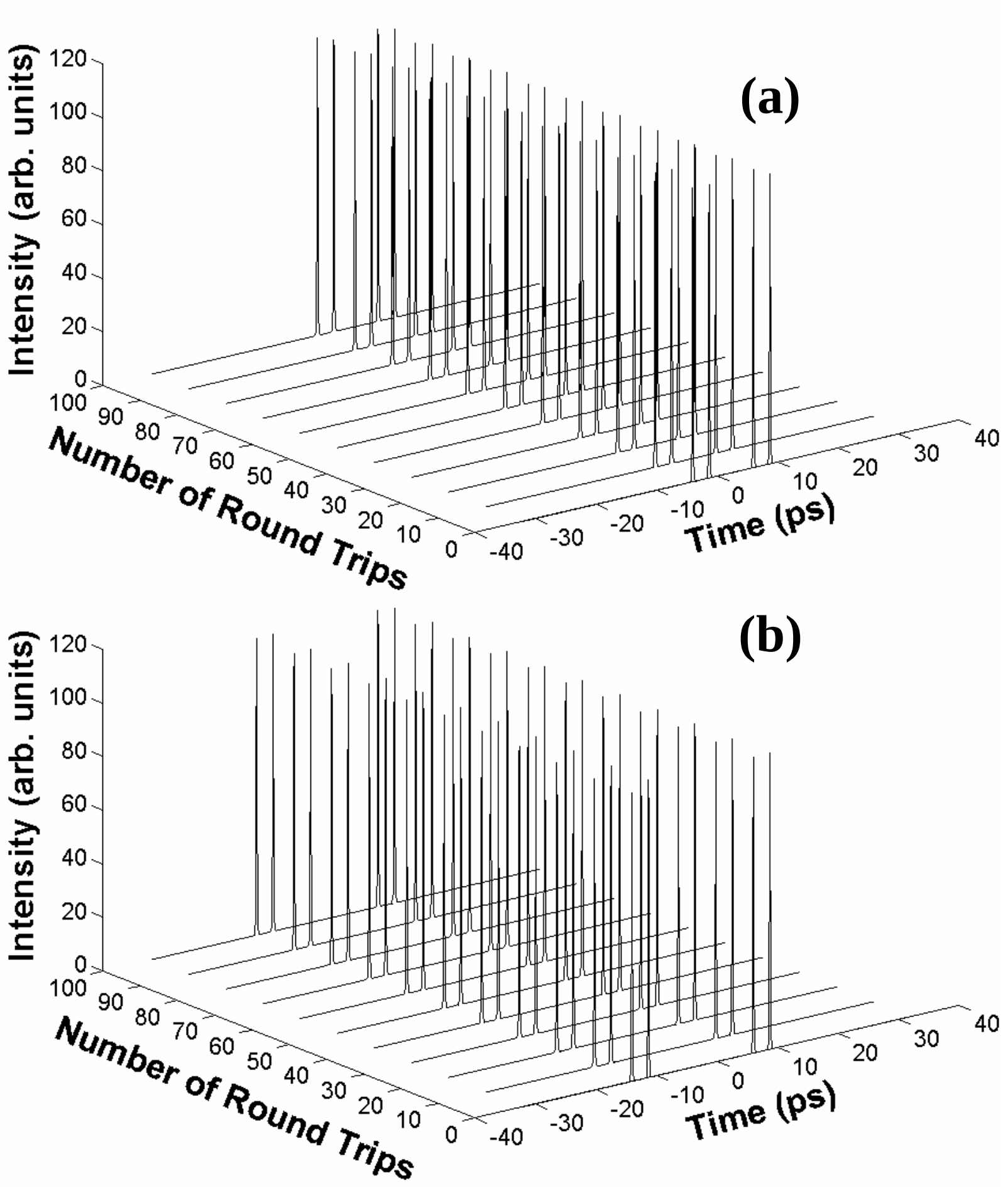

Fig. 9 L. M. Zhao et al. "Bound states of dispersion-managed solitons in a fiber laser with around zero dispersion"